# A METHOD FOR FINDING SIMILAR DOCUMENTS RELYING ON ADDING REPETITION OF SYMBOLS IN LENGTH BASED FILTERING

## HOSSEN AZGOMI[1a] AND MASUMEH GHASEMI MAHSAYEH[b] AND MASOUD MOHAMMADI[c] AND MILAD MORADI RAD[d]

[a]Young Researchers and Elite Club, Rasht Branch, Islamic Azad University, Rasht, Iran
[b]Young Researchers and Elite Club, Rasht Branch, Islamic Azad University, Rasht, Iran
[c]Islamic Azad University, Science and Research, Tehran, Iran
[1d]Department of Computer Engineering, Zanjan Branch, Islamic Azad University, Zanjan, Iran

## ABSTRACT

A basic topic in mining of massive dataset is finding similar items. As an example, finding similar documents can be recommended. In this case many methods are existed. For example, Shingling method and length based filtering are one of them. In Shingling method, from each document, substrings have been selected with symbol name and, they are placed on one set. For finding similar documents, the similarities of sets that related with them have been calculated. In Length based filtering just documents which close these lengths have been compared. These methods don't consider repetition of symbols. With considering the repetition can calculate length of documents with more accurately. In this paper we suggested a method for finding similar documents with considering the repetition of symbols. This method separated documents to better form. The main goal of this paper is presentation a method for finding similar documents with take fewer comparisons and time indeed.

With advances in data collection techniques and extended storage media volume of stored data is growing constantly. As an example, all pages stored in internet can be noted. With the increasing volume of data stored, it is not possible retrieving them with conventional techniques easily. Therefore methods for processing massive data are required. For example finding similar documents in massive dataset of documents is one of these processes (Mitra and et al ; 2002).

In this paper, we first investigated issue of similarity. Then past solutions are listed. In continue presented the proposed method that relying on adding repetition of symbols. Afterwards examined the proposed method practically and its results are expressed. Finally conclusions and future works are given.

## SIMILARITY

First, we should select a criterion for similarity between documents and calculate similarity of document based its. This similarity can dependent on intersection between documents. Namely if the value of intersection is more, then value of similarity is increased. For this purpose, we select the Jaccard similarity as criterion for similarity between documents that based on intersection of sets.

## JACCARD SIMILARITY

The Jaccard similarity of two sets is equivalent to the proportion of the sets' intersection to their union. The Jaccard similarity of the two sets T, S is shown as SIM (S, T) and is calculated like the relationship (1) (Bank and Cole ; 2008):

$$SIM(S,T) = \frac{S \cap T}{S \cup T} \qquad (1)$$

For instance, the Jaccard similarity for the sets {a, b, c} and {a, b, d} is $\frac{2}{4} = 0.5$. One group of issues which the Jaccard similarity covers well is the finding of similar textual documents in a big set (Rajaraman and Ullman ; 2012). Finding the similarity of the texts has different applications that one of them is finding locating scientific and literary plagiarism.

## PAST SOLUTIONS

For calculate the Jaccard similarity between two documents is required that each document converted to a set. It does perform in the Shingling method. In continue, first surveyed the Shingling method. Then length based method is expressed for prevention of the compare of all available documents.

## SHINGLING METHOD

In this method, the documents are displayed in the form of sets. It means that sets are created using the documents. These sets include short strings which are selected from the documents. These short strings are called shingles. It means that shingles are selected of each document, and will be placed in the set related to that document. Using the Jaccard similarity of these sets, the amount of their similarity can be calculated (Manber , 1994).

As we know, a document is a string consisted of characters. A $k$-shingle for a document equals every sub-string with the length of $k$ that has appeared inside that document. In this method, we select a set of $k$-shingles which are repeated once or more and allocate it to that document (Rajaraman and

---

[1]Corresponding author

Ullman , 2012). For instance, suppose that the document D contains the string abcdabd and the value of *k* is 2. Then a set of 2-shingles for the document D will be in line with the set {ab, bc, cd, da, bd}. Please note that that sub-set "ab" has appeared inside the document D two times, but is taken into consideration as a shingle only once.

## LENGTH BASED FILTERING

In the shingling method, in order to find the similar documents, the Jaccard similarity of all the pairs of documents should be calculated and then the similar documents should be obtained. It's obvious that such a task needs a great amount of time. Even. So, we must to avoid investigating all the pairs of documents and review those which are similar with a higher probability. It means that we have to filter the documents in a way or another. To implement filtering, each set obtained from shingling method can be displayed in the form of a string. To display a set, one can first sort the items of the universal set (the union of all sets) based on a certain order. Then the set be displayed as a list of items. These items are sorted based on the order of the items of the universal set. The relevant list is a string of characters which are items of the universal set (Rajaraman and Ullman ; 2012). Here, each items of the universal set is named a symbol. It means that each shingle obtained from document is a symbol.

The easiest way to employ the string presented is sorting the strings based on their length. Then, each "s" string will be compared to the "t" strings which appear after "s" in this list and sequence and are not so lengthy. Suppose that the upper endpoint in the Jaccard distance between the two strings equals J. The Jaccard distance of two sets equals one minus the Jaccard similarity of those sets. The length of each x string will be shown as $L_x$. Please note that $L_s \leq L_t$. The intersection of the intervals shown by the "s" and "t" strings cannot be more than the number of $L_s$ members, while their union has at least as many members as the number of $L_t$ members. As a result, the Jaccard similarity of the "s" and "t" strings which is shown by SIM (s, t) is at most $L_s / L_t$. It means that in order to compare "s" and "t", $J \leq L_s / L_t$ or its equivalent $L_t \leq L_s /J$ (Chaudhuri and et al ; 2006).Suppose that "s" is a string with the length of 9 and we are looking for strings with a Jaccard similarity of at least 0.9. Then "s" will be compared with the strings which are sorted in a sequence based on length and have a length of at most 9 / 0.9 = 10. It means that "s" will be compared to the strings with a length of 9 which are put in order after "s" and all the strings which have a length of 10. It's not needed that "s" be compared to other strings.

## THE PROPOSED METHOD

In Shingling method substrings that obtained from documents were placed in the set. We don't have repetition in the sets. Therefore repetition of shingles doesn't consider in calculating similar documents. For instance in length based filtering may a shingle is repeated twenty times, but only adds one unit to lenght. Namely length of document is shorter nineteen units. It's obvious in this case the document compared to other documents that doesn't have similarity with them. Since in this case calculated length with low accuracy. To solve this problem must first calculate repetition of shingles in Shingling method. For this purpose, we can consider a repetition factor for each shingle and placed repetition of shingle in document to it. Namely shingles that obtain from document have a field that defines its repetition in documents.

With considering repetition of shingles can better be calculated length of strings that obtained from document in lenght based filtering. For this purpose, we can calculate sum of the all repetition factor of the symbols from a document and consider as lenght of string. However for increasing the length accuracy we can multiply each position of symbol to its repetition factor and adding the results together. The length accuracy calculated in this case is more. Since it consider repetition of symbols with this position. Of course the resulting value is not indicating length of string, but it is suitable for separating the documents for decreasing number of comparison. Namely with considering this length in length base filtering should be reduced number of required comparison for finding similar documents. As a result, similar documents are found in less time.

According to the above description the length of each string calculated with relationship $\sum_{i=1}^{L_s} i \times Fs(i)$. In this relationship Fs(i) is define the repetition of each symbol in "s" document. After calculating the length of all documents and sorting those according this value should compared "s" document only with "t" documents that come after it in order and relationship (2) between them is true:

$$\sum_{i=1}^{L_t} i \times Ft(i) \leq \sum_{i=1}^{L_s} i \times Fs(i) \Big/ J \qquad (2)$$

The purpose of comparing two documents is calculating the Jaccard similarity between them. In this method we expect that found similar documents with less number of comparison and therefore in less time than previous method.

## EVALUATION AND PRACTICAL RESULTS

For checking proposed method first we were implemented this method and without repetition method. Then selected a dataset and run these two methods on documents in it. Finally it is expressed summary of the results of the evaluation.

## IMPLEMENTATION

Now we explain details of implementation of proposed method. The implemented algorithm find similar documents in a dataset relying on adding repetition of symbols in length based filtering. The algorithm of proposed method with name Length_Repetition as pseudo code is shown in Figure 1.

```
Length_Repetition(Dataset, k, J)
   Create LRTable(doc, length);
   For each file in Dataset
      S := Load(file);
      RShingling(S, k, Ls1, Ls2);
      For i=0 to Count(Ls1) - 1
         Sh += Ls1.Item(i) + NewLine;
         LR += (i+1) * Ls2.Item(i);
      Save(Sh) in Shingle Directory;
      Add row(file, LR) to LRTable;
   Sort(LRTable) by length;
   Create SimTable(doc1, doc2, Similarity);
   For i=0 to Rowcount(LRTable) – 2
      L := LRTable.Column(length).Row(i) / J;
      m := i + 1;
      While(True)
         If LRTable.Column(length).Row(m) <= L
   then
            d1 := LRTable.Column(doc).Row(i);
            d2 := LRTable.Column(doc).Row(m);
            Sim = Jaccard(d1, d2);
            If Sim >= 0.9 then
               Add row(d1, d2, Sim) to SimTable;
            m += 1;
            If m > Rowcount(LRTable)-1 then
               Exit While;
         Else
            Exit While;
   Save SimTable;
```

**Figure 1: The pseudo code of proposed algorithm**

This algorithm finds similar documents based on proposed method. In this algorithm first calculated symbols and its repetition in each document by the help of RShingling( ) function. Then calculated the length of each document based on repetition of symbols and its positions. In continue for all available documents be determined that which documents must be compared with each document. Jaccard( ) function calculate the value of similarity between two documents. Nevertheless, specified all pairs of similar documents after running algorithm.

## EVALUATION

For practical evaluation, the proposed method and without repetition method was implemented by using the VB.Net programming language. Also NSF Research Awards Abstracts dataset is selected for checking the implemented algorithms. In this dataset, the abstracts of papers at the NSF institute were collected from 1990 to 2003. All documents in dataset are text files. We implemented the algorithms on the 2003 documents. The number of the documents is 645. Also considered k=5 and h=0.9.

First without repetition algorithm is run on the documents. This algorithm is identified 243 pairs of document as similar in the time of 02:11:35. This algorithm compares 25854 pairs of document for finding similar documents. Namely it is dismissed number of 181836 pairs of document from $\binom{645}{2} = 207690$ available pairs of document.

After without repetition algorithm, algorithm of proposed method is run on the documents. This algorithm is also identified 243 pairs of document as similar documents. Its work performed at the time of 01:01:52. This algorithm dismissed number of 193366 pairs of document. Namely 12324 comparisons are performed for finding similar documents. The results of running these two algorithms are given in the Table 1.

**Table 1: The results from running of algorithms**

| Method | Time | Number of comparisons | Number of similar documents |
|---|---|---|---|
| without repetition method | 02:11:35 | 25854 | 243 |
| proposed method | 01:01:52 | 12324 | 243 |

As it can be seen, proposed method do less comparison than without repetition method for finding similar documents. As a result, it finds similar documents in less time. Similar documents in both methods were identical. It means that proposed method finds same similar documents in without repetition method at the time less than half.

## CONCLUSION

In this article, at the first the Jaccard similarity as a criterion for calculating similarity between documents specified. Then shingling method and length based filtering for finding similar documents are introduced. These methods do not consider repetition of symbols in documents. In proposed method computed the repetition of symbols in documents and it used in calculating length of documents. With this length we can separate documents to better form. In continue, proposed method was implemented and it was evaluated in practical aspect. In practical evaluation, this result was obtained that by adding repetition we can find similar document with less comparison and therefore at less time.

In addition to length based filtering, we can separate documents by using of indexing. Indexing is done based on a suitable feature of strings. As an example, we can do indexing. As an example, we can do indexing based on symbols of each document. Therefore presenting a suitable feature in indexing of documents can be a suitable research ground in this regard.